\documentstyle[12pt]{article}

\begin{document}
\title{\bf {Casimir energy-momentum tensor for the Robin  Surfaces in  de Sitter Spacetime }}
\author{M.R. Setare  \footnote{E-mail: rezakord@ipm.ir}
  \\{Physics Dept. Inst. for Studies in Theo. Physics and
Mathematics(IPM)}\\
{P. O. Box 19395-5531, Tehran, IRAN }}
\date{\small{}}
 \maketitle

\begin{abstract}
The energy-momentum tensor for a massless conformally coupled
scalar field in  de Sitter spacetime in the presence of a couple
curved branes is investigated. We assume that the scalar field
satisfies the Robin boundary condition on the surfaces.  Static
de Sitter space is conformally related to the Rindler space, as a
result we can obtain vacuum expectation values of energy-momentum
tensor for conformally invariant field in static de Sitter space
from the corresponding Rindler counterpart by the conformal
transformation.
 \end{abstract}
% \begin{document}
\newpage
% \vspace*{10mm}

 \section{Introduction}
 de Sitter (dS) spacetime is the maximally symmetric solution of Einsten's equation
with a positive cosmological constant. Recent astronomical
observations of supernovae and cosmic microwave background
\cite{Ries98} indicate that the universe is accelerating
 and can be well approximated by a world with a positive cosmological constant. If the
  universe would accelerate indefinitely, the standard cosmology leads to an asymptotic dS
   universe. de Sitter spacetime plays an important role in the inflationary scenario,
   where an exponentially expanding approximately dS spacetime is employed to solve
    a number of problems in standard cosmology. The quantum field theory on dS spacetime
    is also of considerable interest. In particular, the inhomogeneities generated by
    fluctuations of a quantum field during inflation provide an attractive mechanism for
     the structure formation in the universe. Another motivation for investigations of
     dS based quantum theories is related to the recently proposed holographic duality between
     quantum gravity on dS spacetime and a quantum field theory living on boundary identified
     with the timelike infinity of dS spacetime \cite{Stro01}.\\
The Casimir effect is regarded as one of the most striking
manifestation of vacuum fluctuations in quantum field theory. The
presence of reflecting boundaries alters the zero-point modes of a
quantized field, and results in the shifts in the vacuum
expectation values of quantities quadratic in the field, such as
the energy density and stresses. In particular, vacuum forces
arise acting on constraining boundaries. The particular features
of these forces depend on the nature of the quantum field, the
type of spacetime manifold and its dimensionality, the boundary
geometries and the specific boundary conditions imposed on the
field. Since the original work by Casimir in 1948 \cite{Casi48}
many theoretical and experimental works have been done on this
problem (see, e.g.,
\cite{Most97,Plun86,Lamo99,Bord01,Kirs01,Remeo} and references
therein).
     \\The Casimir effect
can be viewed as a polarization of vacuum by boundary conditions.
Another type of vacuum polarization arises in the case of an
external gravitational fields \cite{Birrell,Grib94}.
   Casimir stress for parallel
   plates in the background of static domain wall in four and two dimensions
  is calculated in \cite{{set1},{set2}}. Casimir stress on spherical bubbles immersed in different de Sitter
  spaces in- and out-side is calculated in \cite{{set3},{set4}}.  \\
  It is well known that the vacuum
state for an uniformly accelerated observer, the Fulling--Rindler
vacuum \cite{Full73,Full77,Unru76,Boul75, sahrin}, turns out to be
inequivalent to that for an inertial observer, the familiar
Minkowski vacuum. Quantum field theory in accelerated systems
contains many of special features produced by a gravitational
field avoiding some of the difficulties entailed by
renormalization in a curved spacetime. In particular, near the
canonical horizon in the gravitational field, a static spacetime
may be regarded as a Rindler--like spacetime. Rindler space is
conformally related to the de Sitter space and to the
Robertson--Walker space with negative spatial curvature. As a
result the expectation values of the energy--momentum tensor for a
conformally invariant field and for corresponding conformally
transformed boundaries on the de Sitter and Robertson--Walker
backgrounds can be derived from the corresponding Rindler
counterpart by the standard transformation  \cite{Birrell}.
Vacuum expectation values of the energy-momentum tensor for the
conformally coupled Dirichlet and Neumann massless scalar and
electromagnetic fields in four dimensional Rindler spacetime was
considered by Candelas and Deutsch \cite{CandD}.\\
  In this paper the vacuum expectation value of the energy-momentum tensor are investigated
  for a massless conformally coupled scalar field obeying the Robin boundary conditions on two
  curved surfaces in de Sitter spacetime.
Robin type conditions are an extension of Dirichlet and Neumann
boundary conditions and appear in variety of situations, for
example the casimir effect for massless scalar filed with Robin
boundary conditions on two parallel plates in de Sitter spactime
is calculated in \cite{setm}, the Robin type boundary condition in
domain wall formation is investigated in \cite{setd}. In
Ref.\cite{setsahr} the vacuum expectation value of the surface
energy-momentum tensor is evaluated for a massless scalar field
obeying a Robin boundary condition on an infinite plane moving by
uniform proper acceleration through Fulling-Rindler vacuum. By
using the conformal relation between the Rindler and de Sitter
spacetime, in Ref.\cite{setsahr1} the vacuum energy-momentum
tensor for a scalar field is evaluated in de Sitter spacetime in
presence of a curved brane on which the field obeys the Robin
boundary condition with coordinate dependent coefficients. Robin
boundary conditions naturally arise for scalar and fermion bulk
fields in the Randall-Sundrum model \cite{fla, set6, sahari}.

\section{Vacuum expectation values for the energy-momentum tensor}
We will consider a conformally coupled massless scalar field $%
\varphi (x)$ satisfying the following equation
\begin{equation}
\left( \nabla _{\mu }\nabla ^{\mu }+\frac{1}{6} R\right) \varphi
(x)=0, \label{fieldeq}
\end{equation}
on the background of de Sitter space-time in static coordinates.
In Eq. (\ref{fieldeq}) $\nabla _{\mu }$ is the operator of the
covariant derivative, and $R$ is the Ricci scalar for the de
Sitter space.
\begin{equation}
R=\frac{12}{\alpha^{2}}.  \label{Riccisc}
\end{equation}
The static form of de Sitter space time which is conformally
related to the Rindler space time is given by  \cite{Birrell}
\begin{equation}\label{eqds}
ds_{{\rm
dS}}^{2}=[1-(\frac{r^{2}}{\alpha^{2}})]dt^{2}-[1-(\frac{r^{2}}{\alpha^{2}})]^{-1}dr^{2}-r^{2}
(d\theta^{2}+\sin^{2} \theta d\phi^{2}),
\end{equation}
 To make maximum use of the flat
spacetime calculations, first of all let us present the dS line
element in the form conformally related to the Rindler metric.
With this aim we make the coordinate transformation $x^i\to
x'^{i}=(\tau ,\xi ,y  ,z )$ \cite{setsahr1}

Let us denote the energy--momentum tensor in coordinates $(\tau
,\xi , y, z)$ as $T'_{\mu \nu }$, and the same tensor in
coordinates $(t ,r, \theta ,\phi ) $ as $T_{\mu \nu }$. The
tensor $T_{\mu \nu }$ has the structure
\begin{equation}\label{Tmunu}
T_{\mu }^{\nu }={\mathrm{diag}}(\varepsilon ,-p,-p_{\perp
},p_{\perp })  .
\end{equation}
Our main interest in the present paper is to investigate the
vacuum
expectation values (VEV's) of the energy--momentum tensor (EMT) for the field $%
\varphi (x)$ in the background of the above de Sitter space time
induced by two curved surfaces. We will let the surfaces $\xi=a$
and $\xi=b$, $b>a$ represent the trajectories of these boundaries
for corresponding problem in Rindler spacetime. We will consider
the case of a scalar field satisfying Robin boundary condition on
the surface of the plates, therefore in the Rindler space we have
\begin{equation}
(A_{R}^{(j)}+B_{R}^{(j)}n^{l}_{R_{j}}\nabla_{l}
)\varphi|_{\xi=j}=0,\hspace{1cm}j=a,b
\hspace{1cm}n^{l}_{R_{a}}=\delta^{l}_{1}\hspace{1cm}n^{l}_{R_{b}}=-\delta^{l}_{2}
\label{Dboundcond}
\end{equation}

with constant coefficients $A_{R}^{(j)}$ and $B_{R}^{(j)}$. In
coordinates $x^{i}$ the boundary $\xi=j$, $j=a,b$ are presented
by the surfaces
 \begin{equation}
 \sqrt{\alpha^{2}-r^{2}}=j(1-\frac{r}{\alpha}\cos\theta),
 \label{sur}
 \end{equation}
 in dS spacetime. The boundary condition (\ref{Dboundcond}) in de
 Sitter space is as following
\begin{equation}
(A^{(j)}+B^{(j)}n_{j}^{l}\nabla_{l} )\varphi|_{S_{j}}=0,
\hspace{1cm}j=a,b \label{Dboundcond1}
\end{equation}
where ${S_{j}}$ is given by the surfaces (\ref{sur}).
\\ The presence of
boundaries modifies the spectrum of the zero--point fluctuations
compared to the case without boundaries. This results in the
shift in the VEV's of the physical quantities, such as vacuum
energy density and stresses. This is the well known Casimir
effect. It can be shown that for a conformally coupled scalar by
using field equation (\ref{fieldeq}) the expression for the
energy--momentum tensor can be presented in the form
\begin{equation}
T_{\mu \nu }=\nabla _{\mu }\varphi \nabla _{\nu }\varphi -\frac{1}{6} \left[ \frac{%
g_{\mu \nu }}{2}\nabla _{\rho }\nabla ^{\rho }+\nabla _{\mu
}\nabla _{\nu }+R_{\mu \nu }\right] \varphi ^{2},  \label{EMT1}
\end{equation}
where $R_{\mu \nu }$ is the Ricci tensor. The quantization of a
scalar filed on background of metric Eq.(3) is standard. Let
$\{\varphi _{\alpha }(x),\varphi _{\alpha }^{\ast }(x)\}$ be a
complete set of orthonormalized positive and negative frequency
solutions to the field equation (\ref {fieldeq}), obying boundary
condition (\ref{Dboundcond1}). By expanding the field operator
over these eigenfunctions, using the standard commutation rules
and the definition of the vacuum state for the vacuum expectation
values of the energy-momentum tensor one obtains
\begin{equation}
\langle 0|T_{\mu \nu }(x)|0\rangle =\sum_{\alpha }T_{\mu \nu }\{\varphi {%
_{\alpha },\varphi _{\alpha }^{\ast }\}},  \label{emtvev1}
\end{equation}
where $|0\rangle $ is the amplitude for the corresponding vacuum
state, and the bilinear form $T_{\mu \nu }\{{\varphi ,\psi \}}$ on
the right is determined by the classical energy-momentum tensor
(\ref{EMT1}). Instead of evaluating Eq. (\ref{emtvev1}) directly
on background of the curved metric, the vacuum expectation values
can be obtained from the corresponding Rindler space time results
for a scalar field $\bar{\varphi}$ by using the conformal
properties of the problem under consideration. Under the
conformal transformation $g_{\mu \nu }=\Omega ^{2}\bar{g}_{\mu \nu }$ the $%
\bar{\varphi}$ field will change by the rule
\begin{equation}
\varphi (x)=\Omega ^{-1}\bar{\varphi}(x),  \label{phicontr}
\end{equation}
where for metric Eq.(3) the conformal factor is given by $\Omega
=\frac{\sqrt{\alpha^{2}-r^{2}}}{\xi}$. Now by comparing boundary
conditions (\ref{Dboundcond}) and (\ref{Dboundcond1})   and taking
into account Eq.(\ref{phicontr}), one obtains the relation
between the coefficients in the boundary conditions:
\begin{equation}\label{relcoef}
A^{j}=\frac{1}{\Omega }\left( A_{R}^{j}+B_{R}^{j} n_{j}^{l}\nabla
_{l} \Omega \right) , \quad B^{j}=B_{R}^{j}, \quad x\in S_{j}.
\end{equation}
To evaluate the expression $n_{j}^{l}\nabla _{l} \Omega $ we need
the components of the normal to $S_{j}$ in coordinates $x^i$.
They can be found by transforming the components
$n'^{l}_{a}=\frac{\delta^{l}_{1}}{\Omega}$ and
$n'^{l}_{b}=\frac{-\delta^{l}_{2}}{\Omega}$
 in coordinates
$x'^{i}$:
\begin{equation}\label{normal}
n^{l}_{j}=\left( 0,\pm\frac{j}{\alpha }(\cos \theta -r/\alpha ),
\mp\frac{j}{\alpha r }\sin \theta ,0 \right) ,
\end{equation}
where upper sign refer to $j=a$ and lower sign refer to $j=b$.
Now it can be easily seen that $n^{l}\nabla _{l} \Omega
=\mp\sqrt{\alpha ^2-r^2}/\alpha ^2$ and, hence, the relation
between the Robin coefficients in the Rindler and dS problems
takes the form
\begin{equation}\label{relcoef1}
A^{j}=\frac{jA_{R}^{j}}{\sqrt{\alpha
^2-r^2}}\mp\frac{jB_{R}^{j}}{\alpha ^2}, \quad B^{j}=B_{R}^{j}.
\end{equation}

 The Casimir effect with
boundary conditions (\ref{Dboundcond}) on two parallel plates on
background of the Rindler spacetime is investigated in Ref.
\cite{sah} for a scalar field with a Robin boundary condition.
The boundary part of vacuum expectation value of energy-momentum
tensor is as
\begin{equation}
<0\mid T ^{k}_{i}\mid0>^{B}=\sum_{j=a,b}< T ^{k}_{i}>^{j}+< T
^{k}_{i}>^{ab} \label{bemt}
\end{equation}
where (no summation over i) $< T ^{k}_{i}>^{ab}$ is the
interference term, and $< T ^{k}_{i}>^{j}$ are induced by the
presence of a single plane boundaries located at $\xi=a$ and
$\xi=b$ in the regions  $\xi>a$ and  $\xi<b$. The VEV for the
Fulling-Rindler vacuum without boundaries are given by  $\langle
0_R|T_{i}^{k}
  |0_R\rangle$, All divergences are contained in this purely Fuling-Rindler part.
 These divergences can be regularized subtracting the
corresponding VEV's for the Minkowskian vacuum. Using
Eq.(\ref{EMT1}), for the corresponding difference between the
VEV's of the EMT we have \cite{{CandD},{Cand78}}
\begin{equation}\label{subRind}
  \langle T_{i}^{k}\rangle _{{\mathrm{sub}}}^{(R)}=\langle 0_R|T_{i}^{k}
  |0_R\rangle -\langle 0_M|T_{i}^{k}|0_M\rangle=\frac{-1}{480 \pi^{2}\xi^{4}}
diag(1,-1/3,-1/3,-1/3).
\end{equation}

 The vacuum energy-momentum tensor on static de Sitter space  Eq.(3) is
obtained by the standard transformation law between conformally
related problems (see, for instance, \cite{Birrell}) and has the
form
\begin{equation}
\langle T_{\nu }^{\mu }\left[ g_{\alpha \beta }\right]
\rangle=\xi^{4}( \alpha^{2}-r^{2})^{-2}\langle T_{\nu }^{\mu
}\left[ g_{\alpha \beta }\right] \rangle _{{\rm
Rindler}}+\frac{1}{960\pi^{2}\alpha^{4}}\delta^{\mu}_{\nu}
\label{emtcurved2}
\end{equation}
The expression on the right hand side of the above formula is the
stress-energy tensor for de Sitter space without boundaries. The
vacuum energy-momentum tensor (\ref{emtcurved2}) in the presence
of boundaries has the following form
\begin{equation}
\langle T_{\nu }^{\mu }\left[ g_{\alpha \beta }\right] \rangle _{{\rm ren}%
}=\langle T_{\nu }^{\mu }\left[ g_{\alpha \beta }\right] \rangle _{{\rm ren}%
}^{(0)}+\langle T_{\nu }^{\mu }\left[ g_{\alpha \beta }\right] \rangle _{%
{\rm ren}}^{(B)}.  \label{emtcurved1}
\end{equation}
Where the first term on the right is the vacuum energy--momentum
tensor for the situation without boundaries (gravitational part),
and the second one is due to the presence of boundaries. By taking
into account Eqs.(\ref{subRind},\ref{emtcurved2}) the first term
in Eq.(\ref{emtcurved1})can be rewritten as following
\begin{equation}
\langle T_{\nu }^{\mu }\left[ g_{\alpha \beta }\right] \rangle _{{\rm ren}%
}^{(0)}=\frac{-1}{480 \pi^{2}}(\alpha^{2}-r^{2})^{-2}
diag(1,-1/3,-1/3,-1/3)+\frac{1}{960\pi^{2}\alpha^{4}}\delta^{\mu}_{\nu}.
  \label{emtcurved11}
\end{equation}
The boundary part in Eq.(\ref{emtcurved1}) is related to the
corresponding Rindler spacetime boundary part Eq.(\ref{bemt}) by
the relation
\begin{equation}
\langle T_{\nu }^{\mu }\left[ g_{\alpha \beta }\right] \rangle _{%
{\rm ren}}^{(B)}=\xi^{4}( \alpha^{2}-r^{2})^{-2}(\sum_{j=a,b}< T
^{k}_{i}>^{j}+< T ^{k}_{i}>^{ab}). \label{emtcurved12}
\end{equation}
Now we turn to the interaction forces between the surfaces. The
vacuum force acting per unit surface of the curved surface at
$S_j$ is determined by the ${}^{1}_{1}$--component of the vacuum
EMT at this point. The gravitational part of the pressure
according to Eq.(\ref{emtcurved11}) is equal to
\begin{equation}\label{grper}
P_{g}=-<T^{1}_{1}>=\frac{-1}{960\pi^{2}\alpha^{4}}-\frac{1}{1440
\pi^{2}}(\alpha^{2}-r^{2})^{-2}.
\end{equation}
The first term is the same from both sides of the surfaces, and
hence leads to zero effective force. The second term is negative,
then this force is attractive always. But for infinitely thin
surfaces the second term also is the same from the both sides of
the boundaries and, hence, give the zero effective force. The
corresponding effective boundary part pressures can be presented
as a sum of two terms
\begin{equation}
p_{B}^{(j)}=p_{B1}^{(j)}+p_{B{\rm (int)}}^{(j)},\quad j=a,b.
\label{FintD}
\end{equation}
The first term on the right is the pressure for a single plate at $%
\xi =j$ when the second plate is absent. This term is divergent
due to the well known surface divergences in the subtracted
VEV's. The second term on the right of Eq. (\ref{FintD}),
\begin{equation}
p_{B{\rm (int)}}^{(j)}=-\xi^{4}( \alpha^{2}-r^{2})^{-2}[\langle
T_{1}^{1}\rangle ^{(l)}+\langle T_{1}^{1}\rangle ^{(ab)}]_{\xi=j}
\label{pintD}
\end{equation}
with $j,l=a,b$, $l\neq j$, is the pressure induced by the presence
of the second plate, and can be termed as an interaction force. In
dependence of the values for the coefficients in the boundary
conditions, the effective pressures (\ref{pintD}) can be either
positive or negative, leading to repulsive or attractive forces.
For Dirichlet or Neumann boundary conditions on both plates the
interaction forces are always attractive \cite{sahrin}.
 \section{Conclusion}
In the present paper we have investigated the Casimir effect for a
conformally coupled massless scalar field in the presence of a
couple curved branes, on background of the de Sitter spacetimes
which is conformally related to the Rindler spacetime. We have
assumed that the scalar field satisfies Robin boundary condition
on the surfaces. A plane in Rindler spacetime does not correspond
to a plane in de Sitter spacetime, but to a curved surface.
Likewise, Robin boundary condition in Rindler spacetime
corresponds to a different type of boundary condition in de
Sitter spacetime, it has the same form of Robin boundary
condition but instead of a constant, the coefficient $A^{j}$ in
Eq.(7) is a function of the coordinates. The vacuum expectation
values of the energy-momentum tensor are derived from the
corresponding Rindler spacetime results by using the conformal
properties of the problem. As the boundaries are static in the
Rindler coordinates no Rindler quanta are created \cite{sahrin}.
In the region between the surfaces the boundary induced part for
the vacuum energy-momentum tensor is given by
Eq.(\ref{emtcurved12}), and the corresponding vacuum forces
acting per unit surface of the curved surfaces have the form
Eq.(\ref{FintD}). These forces are presented as the sums of two
terms. The first ones correspond to the forces acting on a single
boundary then the second boundary is absent. The vacuum
polarization due to the gravitational field, without any boundary
conditions is given by Eq.(\ref{emtcurved11}), the corresponding
gravitational pressure part has the form Eq.(\ref{grper}), the
first term in this equation is the same from both sides of the
surfaces, and hence leads to zero effective force. The second
gravitational term is negative, then this force is attractive
always, in contrast the effective boundary  pressures
(\ref{pintD}) can be either positive or negative, leading to
repulsive or attractive forces. Therefore may be in some special
cases the attractive gravitational forces can cancel the effective
boundary  pressures. But for infinitely thin surfaces the second
gravitational term also is the same from the both sides of the
boundaries and, hence, give the zero effective force. In this
case only the boundary part remain, where for Dirichlet or Neumann
boundary conditions on both plates the interaction forces are
always attractive \cite{sahrin}. But for Robin boundary
condition, in dependence of the values for the coefficients in
the boundary conditions, the effective pressures can be either
positive or negative, leading to repulsive or attractive forces.

\end{document}